\documentclass[aps,prl,twocolumn,preprintnumbers,superscriptaddress,amsmath,amssymb]{revtex4}
\RequirePackage{lineno}
\usepackage{graphicx}
\usepackage{subfigure}
\usepackage{epsfig}
\usepackage{xcolor}
\usepackage{dcolumn}
\usepackage{bm}
\usepackage{ulem}
\usepackage{color}
\usepackage{amsmath}
\usepackage{amsfonts}
\usepackage{amssymb}
\usepackage{braket}

\begin{document}

\title{Coexistence of extended flat band and Kekul\'e order in Li-intercalated graphene}

\author{Changhua Bao}
\altaffiliation{These authors contributed equally to this work.}
\affiliation{State Key Laboratory of Low-Dimensional Quantum Physics and Department of Physics, Tsinghua University, Beijing 100084, P. R. China}

\author{Hongyun Zhang}
\altaffiliation{These authors contributed equally to this work.}
\affiliation{State Key Laboratory of Low-Dimensional Quantum Physics and Department of Physics, Tsinghua University, Beijing 100084, P. R. China}

\author{Xi Wu}
\affiliation{Shenzhen Geim Graphene Center and Institute of Materials Research, Tsinghua Shenzhen International Graduate School, Tsinghua University, Shenzhen 518055, P. R. China}

\author{Shaohua Zhou}
\affiliation{State Key Laboratory of Low-Dimensional Quantum Physics and Department of Physics, Tsinghua University, Beijing 100084, P. R. China}

\author{Qian Li}
\affiliation{State Key Laboratory of Low-Dimensional Quantum Physics and Department of Physics, Tsinghua University, Beijing 100084, P. R. China}

\author{Pu Yu}
\affiliation{State Key Laboratory of Low-Dimensional Quantum Physics and Department of Physics, Tsinghua University, Beijing 100084, P. R. China}
\affiliation{Frontier Science Center for Quantum Information, Beijing 100084, P. R. China}

\author{Jia Li}
\affiliation{Shenzhen Geim Graphene Center and Institute of Materials Research, Tsinghua Shenzhen International Graduate School, Tsinghua University, Shenzhen 518055, P. R. China}

\author{Wenhui Duan}
\affiliation{State Key Laboratory of Low-Dimensional Quantum Physics and Department of Physics, Tsinghua University, Beijing 100084, P. R. China}
\affiliation{Frontier Science Center for Quantum Information, Beijing 100084, P. R. China}

\author{Shuyun Zhou}
\altaffiliation{Correspondence should be sent to syzhou@mail.tsinghua.edu.cn}
\affiliation{State Key Laboratory of Low-Dimensional Quantum Physics and Department of Physics, Tsinghua University, Beijing 100084, P. R. China}
\affiliation{Frontier Science Center for Quantum Information, Beijing 100084, P. R. China}

\date{\today}

\begin{abstract}

{\bf Doping graphene near the 1/4 filling to shift the extended flat band and van Hove singularity below \textit{E}$_F$ has been highly desirable. Here we report the experimental observation of an extended flat band  below \textit{E}$_F$ in Li-intercalated graphene. Strong electron-phonon interaction is clearly identified by notable kinks in the band dispersion. Moreover, the evolution of the band structure upon Li intercalation shows that the extended flat band and the Kekul\'e order emerge simultaneously.  Our work provides opportunities for investigating flat band related instabilities and its interplay with the Kekul\'e order. }

\end{abstract}

\maketitle

In solids, a flat band with a high density of states (DOS) near  the Fermi energy \textit{E}$_F$ can enhance many-body interactions such as electron-phonon (el-ph) interaction and electron-electron correlation, thereby increasing instabilities toward exotic ordered states.  For example, in magic-angle twisted bilayer graphene (MABLG) \cite{MacDonaldPNAS2011}, the flat band can be induced through hybridization of the Dirac cones via the moir\'e superlattice potential and novel phenomena such as superconductivity \cite{PablotBLG}, Mott insulator \cite{PabloMott} and charge order \cite{AndreiNat2019} have been discovered.  Instead of band hybridization near the Dirac point, a different  strategy for inducing a flat band is doping graphene to near 1/4 filling, so that the extended flat band near the van Hove singularity (VHS, see Fig.~1(a) for schematic band structure) is shifted to \textit{E}$_F$.  Graphene near 1/4 filling has attracted extensive interests with intriguing predictions such as instabilities toward \textit{d}-wave superconductivity by repulsive electron-electron interaction \cite{DoniachPRB2007,Kivelson10,BaskaranPRB2010}, topological superconductivity \cite{ChubukovNP12}, and chiral spin density wave \cite{LeeDH12, RanYPRX14}.

In graphene, the $\pi^*$ band VHS is at $\sim$ 3 eV above the Dirac point, and an extremely high electron doping is required to shift it to \textit{E}$_F$. Experimentally, electron doping of graphene has been achieved by adsorption of alkaline metals \cite{RotenbergPRL2010,Takahashi2011,Takahashi2012,HitosugiSTMPRL2015,Jacobi2016,Gruneis2018} or Gd, Yb \cite{Starke2019PRB2,LanzaraYb2014,Starke2019PRB1,Starke2020PRL}.  In particular, combining Yb intercalation with K adsorption dopes graphene beyond the VHS \cite{Starke2020PRL}, however, the Dirac cone of graphene is severely distorted due to the hybridization with the 4f orbital of Yb. 
In this work, we report the existence of an extended flat band near 1/4 doping in Li-intercalated graphene by angle-resolved photoemission spectroscopy (ARPES) measurements.  Significant band renormalization is observed as a result of strong el-ph interaction. Moreover, by monitoring the evolution of the extended flat band and replica Dirac cone upon Li intercalation, we find that the extended flat band and the Kekul\'e order co-develop, suggesting a strong connection between these two intriguing properties and providing important insights for understanding the underlying mechanism. 

\begin{figure*}[htbp]
	\centering
	\includegraphics[width=16.8 cm]{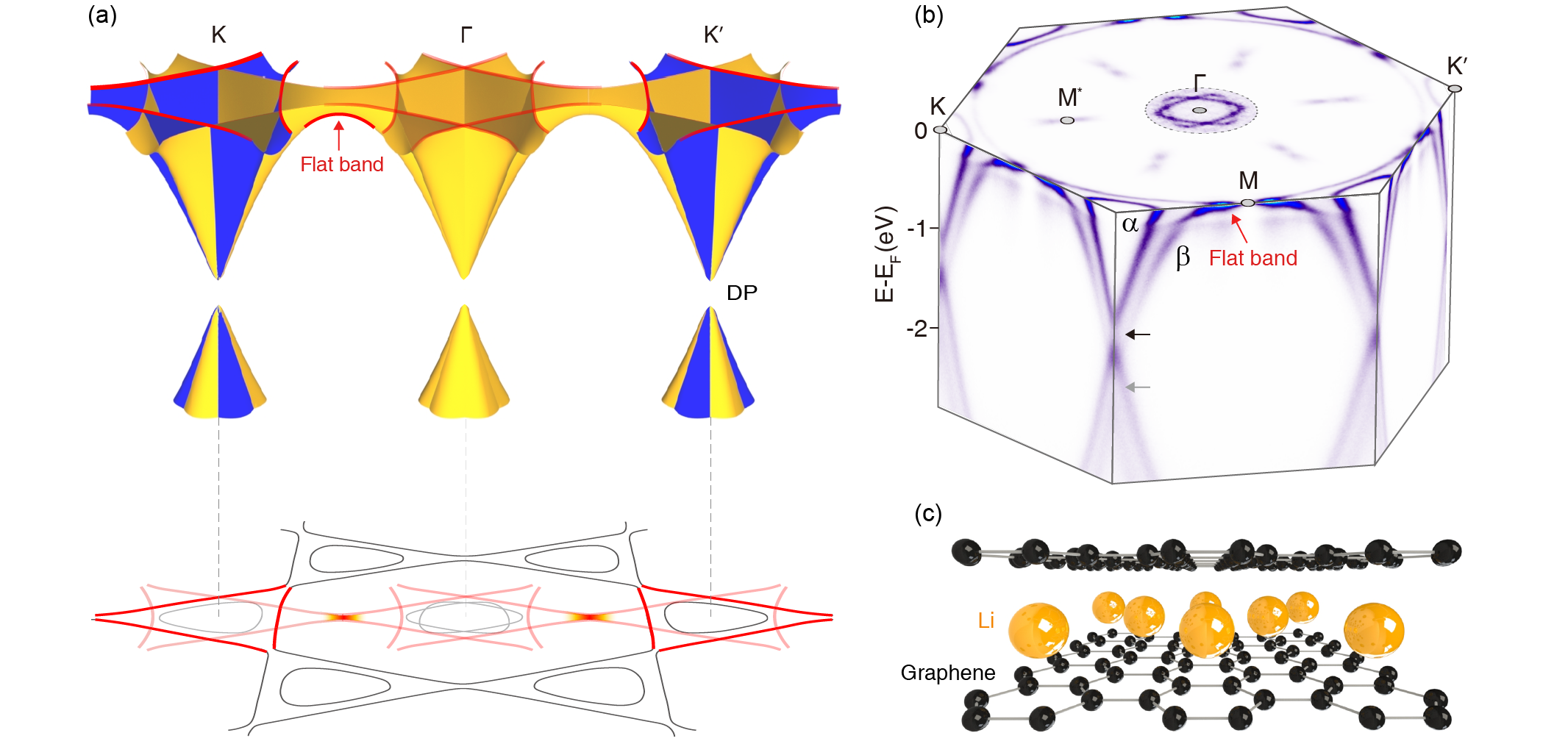}
	\caption{(a) A schematic for the flat band near the van Hove singularity point in Kekul\'e-ordered graphene and constant energy maps of the Dirac cones. (b) The full three-dimensional electronic structure of Kekul\'e-ordered graphene measured by ARPES. The intensity inside the dashed circle around $\Gamma$ is enhanced by a multiplication factor of 15 for better visualization. (c) A schematic drawing of Li-intercalated bilayer graphene.}
\end{figure*}

The Li-intercalated bilayer graphene is obtained by intercalating Li \cite{JohanssonARPES2010,zhou2021} to monolayer graphene on SiC substrate \cite{Xue2013} (see Methods in \cite{Supplementary}). Figure 1(b) shows an overview of the three-dimensional electronic structure from ARPES measurements. The replica pockets around the $\Gamma$ point indicate the Kekul\'e order and chiral symmetry breaking reported recently \cite{zhou2021}.  In this work, we focus on a different aspect, namely, the existence of extended flat band with strong el-ph interaction near 1/4 doping and its connection to the Kekul\'e order. Two large pockets labeled as $\alpha$ and $\beta$ are observed near each Brillouin zone (BZ) corner with the Dirac points at energies of -1.27 and -1.82 eV, indicating a higher doping level than previously reported Li-intercalated graphene \cite{JohanssonARPES2010,Takahashi2012,OttavianoPRB15,Jacobi2016}.  The Dirac point energy of -1.82 eV is at even lower energy than the -1.6 eV in Yb-intercalated graphene \cite{Starke2020PRL}. The absence of hybridization  is in overall agreement with calculated electronic structure \cite{MauriNatPhys2012LiC} and distinguishes it from Ca-intercalated graphene \cite{Takahashi2012,HitosugiSTMPRL2015,MauriNatPhys2012LiC} or Yb-intercalated graphene \cite{Starke2020PRL}.  Such high electron doping is made possible by monitoring the electronic structure with ARPES measurements in real time during Li intercalation, which allows to achieve the optimum doping level (see Fig.~S1-3 and attached movie for more details in \cite{Supplementary}).

\begin{figure*}[htbp]
	\centering
	\includegraphics[width=16.8 cm]{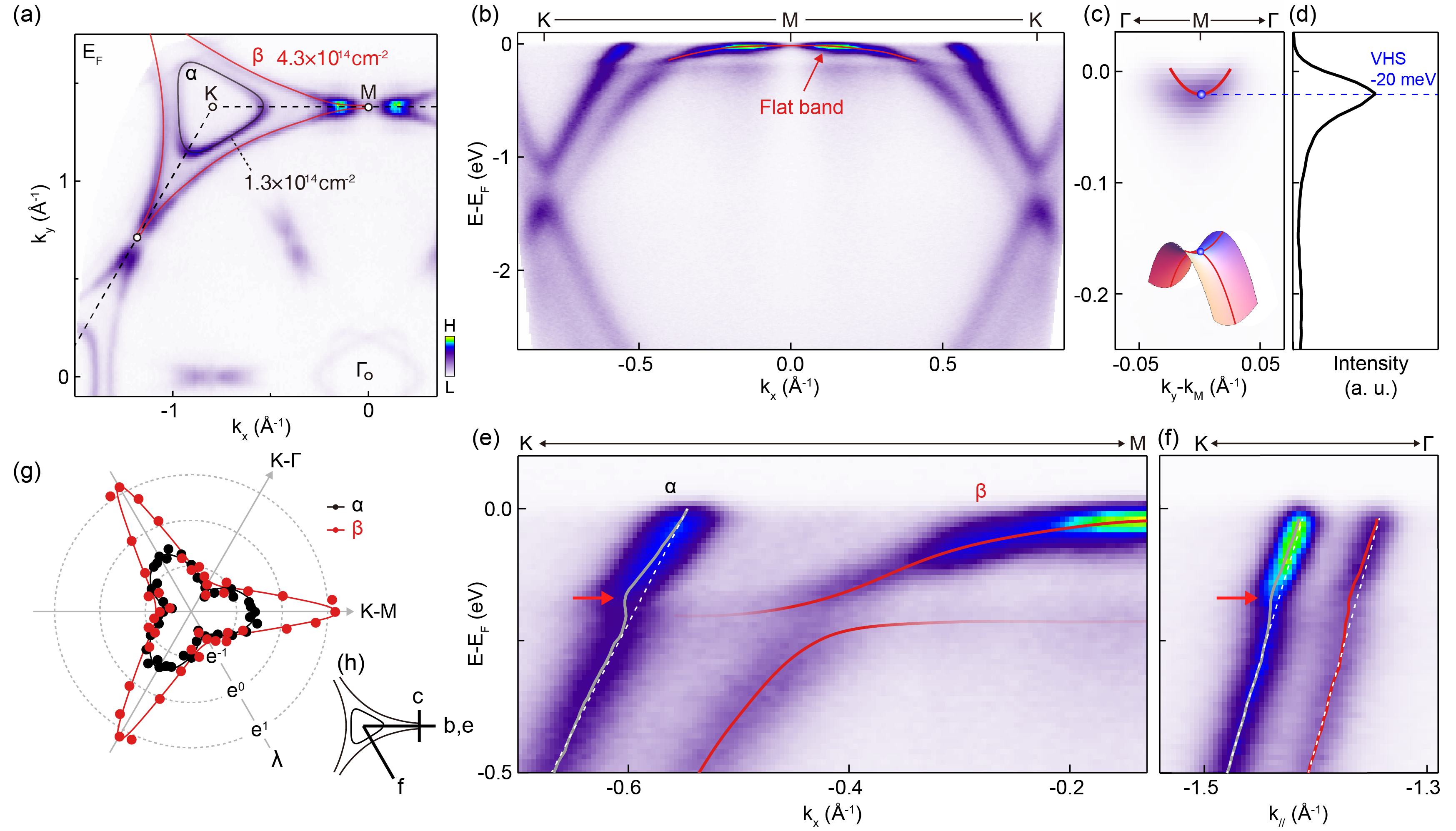}
	\caption{(a) Fermi surface map. (b) Dispersion image along the K-M-K direction. (c) Dispersion image along the $\Gamma$-M-$\Gamma$ direction. The inset is a schematic of the saddle point. (d) EDC at the M point, showing that the VHS (blue dot) is at -20 meV. (e) Dispersion image around \textit{E}$_F$ along the K-M direction. (f) Dispersion image around \textit{E}$_F$ along the K-$\Gamma$ direction. (g) Extracted angular dependence of el-ph coupling constant of two bands. (h) Schematics for directions of corresponding dispersion images.}
\end{figure*}

Figure 2(a) shows the Fermi surface map, from which the carrier concentrations can be directly extracted from the size of the Fermi pockets through the Luttinger theorem \cite{Luttinger1960}. The extracted carrier concentrations are 1.3$\times$10$^{14}$ cm$^{-2}$ and 4.3$\times$10$^{14}$ cm$^{-2}$ for $\alpha$ and $\beta$ pockets respectively, which are equivalent to 0.07 and 0.23 electrons per unit cell.  Even higher carrier concentrations of 1.4$\times$10$^{14}$ cm$^{-2}$ and 5.1$\times$10$^{14}$ cm$^{-2}$, which correspond to 0.08 and 0.27 electrons per unit cell, can be obtained by second Li intercalation of the same sample after annealing (see Methods and Fig.~S4 in \cite{Supplementary}).  The high carrier concentration is in line with the superdense Li intercalation between two graphene sheets with estimated carrier concentration of  4$\times$10$^{14}$ cm$^{-2}$ \cite{SmetLiAmount}, and here we can resolve each electron pocket and extract the corresponding carrier concentration directly.

Figure 2(b) and 2(c) shows dispersion images measured along two high symmetry directions through the M point.
The dispersion is quite flat when approaching the M point and extended in a large momentum range. This corresponds to the top of the valence band along the K-M direction (Fig.~2(b)) and the bottom of the conduction band along the $\Gamma$-M direction (Fig.~2(c)), indicating that it is a saddle point (schematic in the inset of Fig.~2(c)).  Energy distribution curve (EDC) at the M point shows that the saddle point VHS is at -20 meV, thus Li-intercalated graphene provides an important platform for investigating flat band induced instabilities around the 1/4 filling \cite{ChubukovNP12, LeeDH12, RanYPRX14}. Importantly, the doping beyond the VHS and the absence of hybridization or orbital contribution from the intercalated Li, allow to probe the many-body interactions of the intrinsic Dirac fermions.

The high carrier concentration leads to strong el-ph coupling. Figure 2(e) shows a zoom-in of the dispersion image near \textit{E}$_F$.  For the $\alpha$ pocket which does not reach the doping level of the VHS, a renormalization of the dispersion, a kink, is clearly observed at the energy of $-180\pm20$ meV (pointed by the red arrow), while for the $\beta$ band which is beyond the VHS, the el-ph coupling is even stronger, with a strongly deviated dispersion at similar energy. Comparison of this phonon energy with the calculated phonon dispersion for the Kekul\'e-ordered graphene \cite{Zhou2021Tr} shows that the phonons that are coupled with electrons are the in-plane optical phonon A$_{1g}$. The  extracted el-ph coupling strength from the renormalization of the Fermi velocity (see Fig.~S5 for more details in \cite{Supplementary}) is $\lambda_{\alpha}^{KM}=0.55$ along the K-M direction for the $\alpha$ pocket, and $\lambda_{\beta}^{KM}=3.19$ for the $\beta$ pocket.  The el-ph coupling constant is quite anisotropic, and the el-ph coupling strength is reduced to $\lambda_{\alpha}^{K\Gamma}=0.25$ and $\lambda_{\beta}^{K\Gamma}=0.27$ along the K-$\Gamma$ direction (Fig.~2(f)).  The anisotropic ratio increases from 2.2 for the $\alpha$ pocket (black symbols in Fig.~2(g)) to 11.8 for the $\beta$ pocket (red symbols in Fig.~2(g)).
The el-ph coupling constant is larger than previously reported values of $\lambda$ = 0.14 to 0.61 for pristine graphene \cite{lanzara2008}, alkaline metal doped graphene \cite{Eli2007,GruneisNC14}, superconducting Li or Ca doped graphene \cite{MauriNatPhys2012LiC,Damascelli2015} and graphite \cite{VallaPRL2009}, due to the higher carrier concentration here. Such strong el-ph coupling may favor instabilities such as superconductivity in graphene and graphite \cite{VallaPRL2009,VallaPRL2011,ZX2014}. For example, superconductivity with T$_c$ of up to 8.1 K has been predicated in monolayer LiC$_6$ due to the enhanced el-ph coupling \cite{MauriNatPhys2012LiC}. Experimentally, the superconducting gap has been reported in Li-doped graphene \cite{Damascelli2015} and the Meissner effect has been reported in Li-intercalated graphene \cite{LeeHYJPCM2017}.  Here, the observation of stronger el-ph coupling in Li-intercalated graphene provides new possibilities for achieving superconductivity with a higher $T_c$, and future experimental efforts such as transport measurements are needed.

\begin{figure*}[htbp]
	\centering
	\includegraphics[width=16.8 cm]{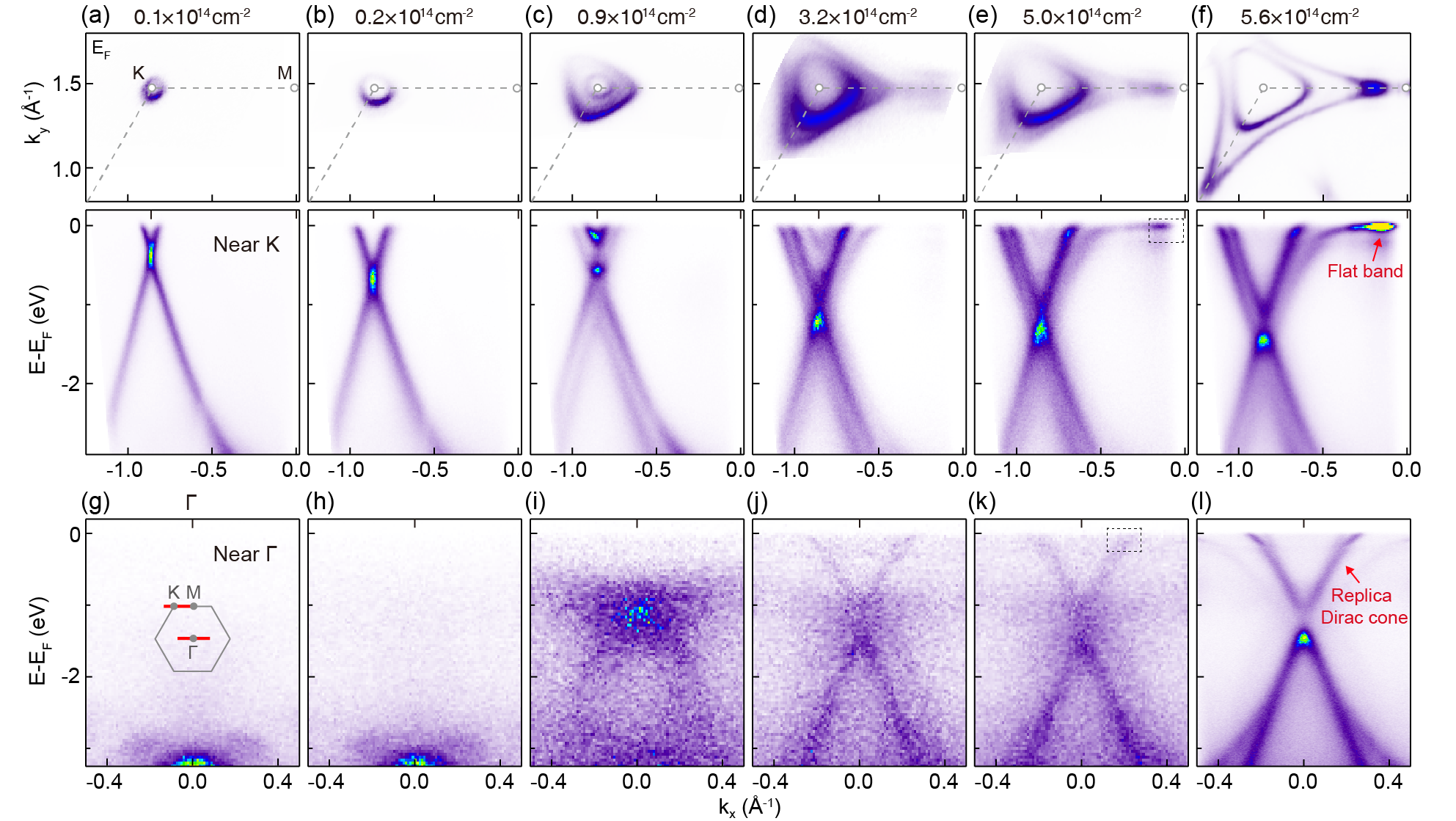}
	\caption{(a-f) Fermi surface maps around the K point and dispersion images along the K-M direction at different doping levels (total carrier concentration). (g-l) Dispersion images along the $\Gamma$-K direction at different doping levels the same as (a-f). The inset in (g) indicates the directions of dispersion images in (a-f) and (g-l).}
\end{figure*}

To further investigate if there is any correlation between the extended flat band and  Kekul\'e order, we show in Fig.~3 the evolution of the electronic structures during the Li intercalation process. Here the size of the Fermi pocket (Fig.~3(a)-(f)) is used to quantify the carrier concentration, while the dispersions near K (middle panels) and near $\Gamma$ (Fig.~3(g)-(l)) are used to monitor the extended flat band and the Kekul\'e order respectively.
Upon Li deposition, the carrier concentration increases, and at carrier concentration of $n=0.9\times10^{14}$ cm$^{-2}$ (see Fig.~3(c)), a new Dirac cone appears, indicating decoupling of the buffer layer from the SiC substrate and the formation of two graphene layers \cite{OttavianoPRB15}. More Li deposition leads to increasing electron doping, and the replica Dirac cone at the $\Gamma$ point becomes detectable at a carrier concentration of $n=3.2\times10^{14}$ cm$^{-2}$ (Fig.~3(j)), indicating the formation of the Kekul\'e order. Further Li deposition leads to a larger $\beta$ pocket (Fig.~3(e)), and the flat band intensity becomes stronger and finally saturated at $n=5.6\times10^{14}$ cm$^{-2}$ (Fig.~3(f)). The emergence of both the extended flat band and replica Dirac cones near $\Gamma$ suggests that they not only coexist, but also are strongly connected. 

\begin{figure*}[htbp]
	\centering
	\includegraphics[width=16.8 cm]{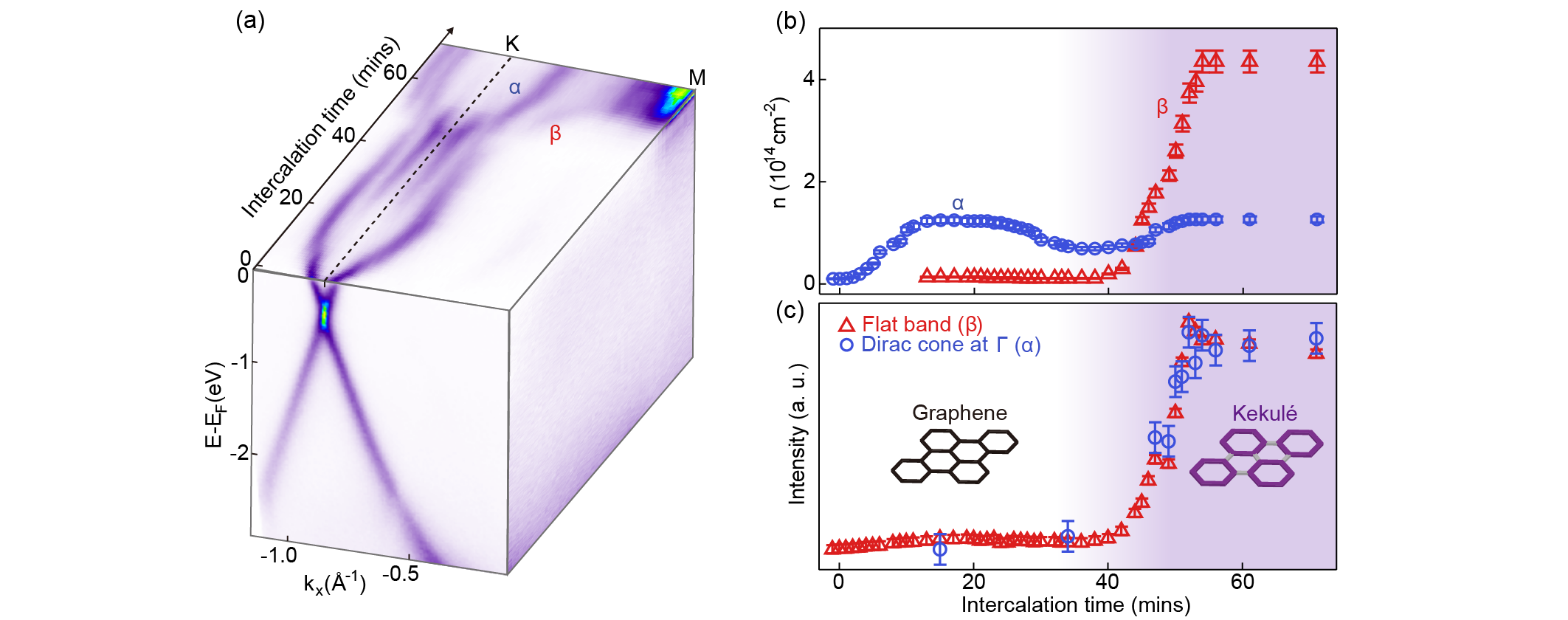}
	\caption{(a) Continuous evolution of the band structures during Li intercalation. (b) Evolution of carrier concentration for two bands $\alpha$ and $\beta$ during Li intercalation. (c) Intensity evolution of flat band and replica Dirac cone during Li intercalation, by integrating the ARPES intensity inside the black dashed boxes in the bottom panel of Fig.~3(e) and Fig.~3(k) respectively. The error bars are estimated from the noise level.}
\end{figure*}

To further confirm the co-development of the extended flat band and the Kekul\'e order, Figure 4(a) shows the continuous evolution of the electronic structure during Li intercalation, and a quantitative analysis of the intensity for the extended flat band and the replica Dirac cone is shown in Fig.~4(c).  A comparison of the temporal evolution of the intensity for the extended flat band (red triangle in Fig.~4(c)) and replica Dirac cone (blue circle) shows that they have the same trend, and they both increase with the carrier concentration of the $\beta$ pocket (red symbol in Fig.~4(b)), thereby confirming the coexistence and co-development of the flat band and the Kekul\'e order. 

To understand the coexistence of the Kekul\'e order and the extended flat band, we first discuss the possible mechanisms of the Kekul\'e order. There are  three main types of mechanisms for the Kekul\'e order: Friedel oscillation mediated adatom interaction, screened Coulomb interaction between adatoms, and adatom-substrate interaction. The Friedel oscillation mediated adatom interaction is the most well-known mechanism for realizing hidden Kekul\'e order in graphene with dilute adatoms \cite{AltSchulerSSC,PasupathyNatPhys}, where the adatoms interact with each other coherently through the Friedel oscillation, and the charge density oscillates with a wavelength $\lambda$ = $\pi/k_F$ ($k_F$ is the Fermi wave vector) to form ordered states. For pristine graphene, $k_F$ is equal to $\Gamma$-K distance and $\lambda$ is equal to the period of Kekul\'e order, thereby favoring the formation of the Kekul\'e order. However, this picture becomes invalid at high doping, because $k_F$ strongly deviates from $\Gamma$-K distance and $\lambda$ is no longer equal to the period of the Kekul\'e order, making it an unlikely mechanism for our Li-intercalated graphene. 

The more likely mechanism for the Kekul\'e order is the direct Coulomb repulsion, complemented by adatom-substrate interaction. The Coulomb repulsion interaction would be screened by graphene, leading to the smaller effective charge and faster decay, however, since the intercalated Li atoms are highly ionized, a stronger Coulomb interaction is expected, increasing the possibility of forming ordered states \cite{Xue2012}.  This, however, is not sufficient to determine the Kekul\'e order, and  adatom-substrate interaction needs to be taken into account, in order to enforce the adatoms to stay in specific positions to maintain the lowest energy, namely the templating effect \cite{Zhang2002,HeLACS2018}. Our first-principles calculations show the hollow site of graphene has the lowest energy (-1.06 eV) for a Li atom than the top (-0.87 eV) or bridge (-0.93 eV) site \cite{zhou2021}, and therefore Li atoms prefer the hollow sites and modify the C-C bonding surrounding them, leading to Kekul\'e-O type bond texture. Therefore, the screened Coulomb interaction and adatom-substrate interaction jointly lead to the formation of the Kekul\'e order in the Li-intercalated graphene.

Along this line, the co-development of the extended flat band and the Kekul\'e order can now be understood as the following. The emergence of the flat band relies on the large charge transfer from Li, while the residual positive charge in the Li ions enhances the screened Coulomb repulsion interaction, thus favoring the formation of Kekul\'e order. In this way, the extended flat band and the Kekul\'e order are intrinsically linked by the charge transfer which contributes to the large doping and the induced Coulomb interaction. This picture is also supported by first-principles calculations (see Fig.~S6 for more details in \cite{Supplementary}).

In summary, we report the coexistence of the extended flat band and the Kekul\'e order in Li-intercalated graphene, thereby identifying an intriguing platform for investigating both electronic instabilities \cite{DoniachPRB2007,Kivelson10,BaskaranPRB2010,ChubukovNP12,LeeDH12, RanYPRX14} and Kekul\'e order related physics \cite{Mudry2007,Naumis2019}. Whether the flat band and the Kekul\'e order can interact with each other is an interesting open question which needs more investigations in the future, in particular, their interplay may lead to new complex ordered phases by coupling the nesting vector \cite{LeeDH12,RanYPRX14} and the Kekul\'e order vector, and intriguing Kekul\'e superconductivity \cite{Herbut2010}.

\begin{acknowledgments}
\section*{ACKNOWLEDGMENTS}
We thank Hong Yao, Congjun Wu and Peizhe Tang for useful discussions. This work was supported by the National Key R\&D Program of China  (Grants No.~2021YFA1400100, 2020YFA0308800),  the National Natural Science Foundation of China (Grant No.~11725418). J. L. was supported by the National Natural Science Foundation of China (Grant No. 11874036) and Local Innovative and Research Teams Project of Guangdong Pearl River Talents Program (2017BT01N111).

\end{acknowledgments}


\end{document}